# Synchronized Wavelength-Swept Signal Transmission and its Ability to Evade Optical Reflection Crosstalk

**Bernhard Schrenk**
*AIT Austrian Institute of Technology,*
*Center for Digital Safety&Security,*
*1210 Vienna, Austria.*
*Author e-mail address: bernhard.schrenk@ait.ac.at*

**Abstract:**
Coherent homodyne detection requires a precise matching of emission wavelengths between transmitter and local oscillator at the receiver. Injection-locking can provide all-optical synchronization of the emission frequencies, even under wavelength-swept emission. By adapting the sweep parameters to the conditions in the optical fiber plant, transmission impairments can be mitigated. In this regard I experimentally demonstrate that a wavelength-hopping yet locked transceiver pair, which builds on conceptually simple externally modulated laser technology, features a much higher robustness to reflection crosstalk. The reception penalty due to distortions arising at a Fresnel reflection in the transmission path can be reduced by >90% at a low optical signal-to-reflection ratio of ~0 dB.

Back-scattering of light along transmission fibers and Fresnel reflections can quickly deteriorate the reception performance in passive optical networks (PON) [1]. Mitigation techniques such as frequency dithering [2], wavelength conversion [3,4] or coding [5,6] have been demonstrated for PONs based on wavelength re-use, while commercial PON systems apply spectral splitting of down- and upstream wavebands to avoid reflection penalties momentarily. A possible migration towards coherent PONs puts this technical challenge back in the spotlight. Although Nyquist filtering of spectrally close down-/upstream signals in heterodyne systems [7] has been proven similarly beneficial as Nyquist pulse shaping for adjacent intradyne downstream channels [8], there has been no thorough investigation on the mitigation of in-band upstream reflection noise in intradyne PONs. Previous research works have proposed narrowband electrical filtering to exclusively suppress optical carrier beat noise in remotely seeded self-homodyne schemes [9,10], which cannot compensate for reflections arising from concurrent data transmission. Moreover, estimation of beat noise for subsequent compensation [11] is hardly applicable and requires digital signal processing resources.

In this letter, I experimentally demonstrate reflection mitigation for conceptually simple coherent links for which all transceivers are based on externally modulated lasers (EML). Synchronous wavelength-swept emission for transmitters and local oscillators (LO) similar as in frequency hopping evades beat noise at the optical level and enables optical signal-to-reflection ratios (OSRR) of ~0 dB at which it recovers more than 90% of reception penalty.

The coherent optical link between central office (CO) and the optical network terminal (ONT) builds on an EML as widely adopted laser device to serve as transmitter and coherent receiver, respectively. Both functions have been demonstrated in a full-duplex coherent transceiver configuration [12,13]. The complexity of the EML-based coherent PON transceiver is comparable to that of systems building on intensity modulation and direct detection (IM/DD) systems. It features an electro-absorption modulator (EAM) as fast PIN photodetector and modulator, and an in-line distributed-feedback (DFB) laser as local oscillator (LO). Through partial injection of the incident signal at $\lambda^*$ the DFB laser at $\lambda$ can be injection-locked. This leads to homodyne reception with precise frequency match of the downlink signal and the LO. Moreover, this all-optical locking method is stable for low injection power [12].

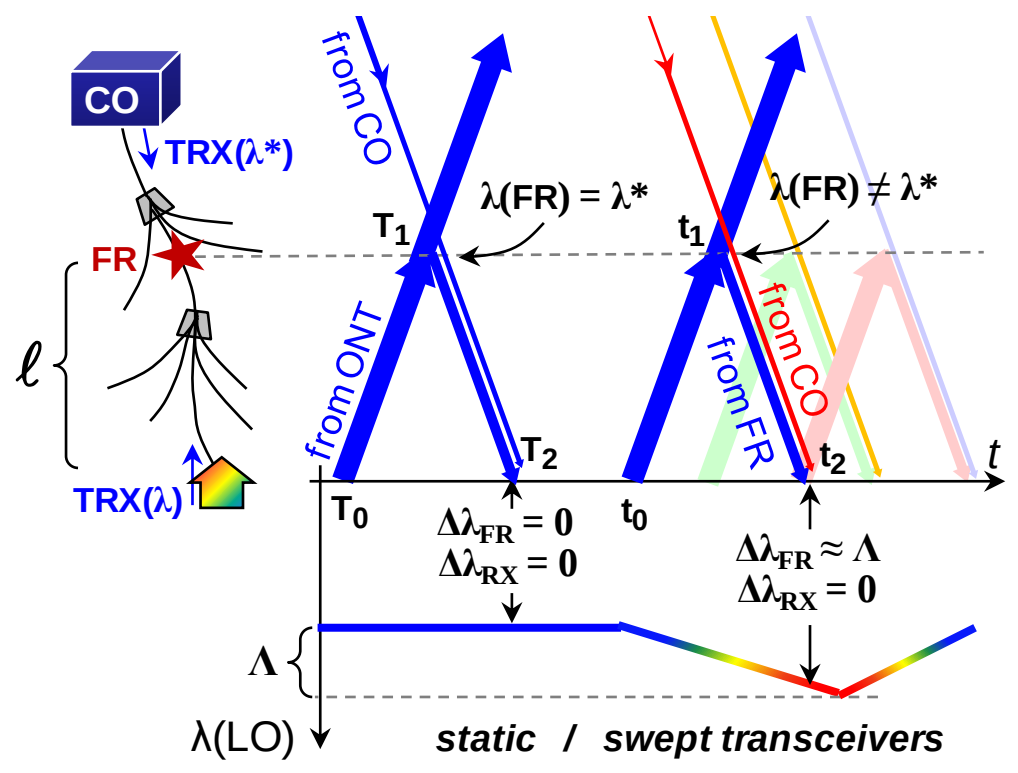

Fig. 1. Frequency swept transmission for mitigation of crosstalk arising from Fresnel reflections at the ODN.

This letter investigates the reflection tolerance of the injection-locking process of this coherent emitter/detector element in presence of a Fresnel reflection (FR). Figure 1 illustrates the path-time diagram for signals transmitted from CO and ONT. The uplink at the LO wavelength λ is precisely locked to the downlink wavelength ($\Delta\lambda_{RX} = 0$) due to coherent homodyne reception. With static optical emission frequencies, corresponding to DC-biased lasers at CO and ONT, the FR causes crosstalk ($T_1$) originating from the upstream. This crosstalk falls in-band with the downlink signal ($\Delta\lambda_{FR} = 0$) and the downlink reception ($T_2$) is penalized. The proposed mitigation scheme jointly sweeps the optical frequencies of CO and ONT transceivers with a peak deviation Λ that is larger than the data bandwidth. The DFB modulation frequency of the sweep sequence is tailored to the reach $\ell$ between ONT and FR. In this way the detuning is maximized to Λ at the round-trip of the reflected light and the crosstalk is shifted ($t_1$) out of the downlink bandwidth at λ*. Since both, downlink seed laser and LO are following the same sequence to ensure the EML at the ONT remains tethered to that at the CO, coherent homodyne reception can be retained ($t_2$: $\lambda_{RX} = 0$); however, with the essential difference that the reflected signal is now spectrally displaced ($\lambda_{FR} = \Lambda$) and out of the reception band. For stable locking of the EML pair, Λ is chosen larger than the locking range of typically 100 MHz [12].

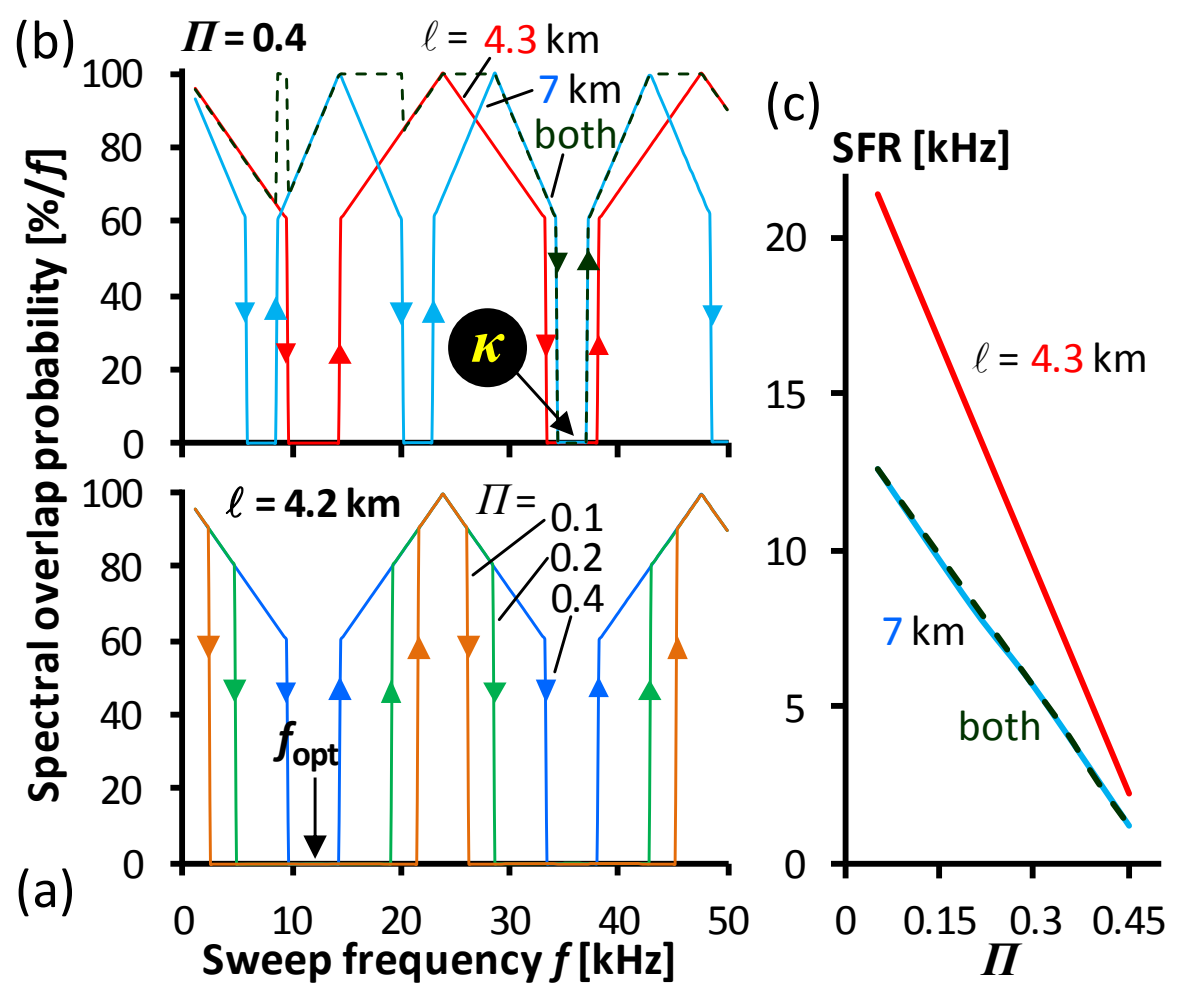

Fig. 2. Overlap probability for crosstalk and signal spectrum under wavelength-swept operation for (a) constant reach to the FR and (b) two FR. (c) Compatible sweep frequency near the optimal setting $f_{opt}$.

In order to estimate the probability of spectral overlap between signal and crosstalk arising at a FR at a certain reach $\ell$, the spectral displacement is analytically elaborated. The aim of the wavelength-swept emission is to maximize the spectral spacing $\Delta\lambda_{FR}$ between the signal and the (delayed) crosstalk at any time.

Let the wavelength sweep of the signal follow a sawtooth function leading to the optical emission frequency $\nu_S(t) = t\Delta F/T_{per}$, where ΔF is the peak deviation of the optical emission frequency sweep and $T_{per}$ the period of the sweep function. The crosstalk of the FR will show a delayed replica with optical frequency $\nu_{FR}(t) = \nu_S(t+\Delta t)$. The delay Δt is governed by the reach $\ell$ to the FR according to the round-trip time, $\Delta t = 2\ell/c$, where $c$ is the group velocity in the optical fiber. The spectral displacement Δν of the crosstalk is given by the difference of optical frequencies impinging on the receiver, $\Delta\nu(t) = \nu_{FR}(t) - \nu_S(t) = 2\ell\Delta F/cT_{per}$. The optimal sweeping frequency is yielded when the displacement is both, constant and maximized to the full swing ΔF. This is the case for $f_{opt} = 1/T_{per} = c/2\ell$.

Crosstalk may arise at multiple FR in the optical distribution network (ODN) and is taken into consideration irrespectively of its actual magnitude in this model. Figure 2a presents the respective spectral overlap probability $f$T* in a sweep period for an ideal sawtooth sequence without falling ramp at a variable sweep frequency $f$. The condition of overlap is defined through the time portion T* within the sweep period for which the optical spectra of signal and back-reflection fall together. The estimation is shown for a reach of 4.3 km to the FR. The parameter $\Pi = f_u/\Delta F$ is defined by the uppermost signal frequency $f_u$ and the frequency deviation applied by the sweep. While $\Pi = 0.5$ marks the border where the swing ΔF is insufficient to sweep beyond the highest signal frequency so that the displacement vanishes entirely and independently of the sweep frequency, the ideal case $\Pi = 0$ corresponds to an infinite frequency swing and crosstalk displacement introduced by the wavelength sweep. Practically the frequency swing ΔF will be chosen according to the modulation bandwidth. For the particular case of an injection-locked EML, the locking range replaces $f_u$ for small signal bandwidths.

For an optimal sweep frequency $f_{opt}$ = 12.1 kHz that is adjusted to the reach of the FR, the wavelength-swept light evades the back-reflection from the FR through a maximum of spectral displacement, as introduced in Fig. 1. This displacement can be retained over the entire sawtooth ramp so that there is actually no overlap with the crosstalk during the entire sweep period. As $\Pi$ increases, corresponding to a relative low frequency deviation ΔF with respect to the signal bandwidth, the tolerable sweep frequency range (SFR) around the optimal setting $f_{opt}$, at which this condition holds, becomes smaller. Yet, for a small frequency deviation ($\Pi$ = 0.4) the range of 4.8 kHz is relatively large. For the later experiment a $\Pi$ of 0.08 applies due to modulation settings. It is effectively advanced to 0.21 for the particular case of an injection locked EML since the locking range is not to be contaminated by carrier crosstalk in order to ensure stable operation.

Fig. 2b shows the dependence of the optimal sweep frequency $f$ for a high $\Pi$ = 0.4. As the round-trip time to the FR increases, as shown for $\ell$ = 7 km, crosstalk is back-reflected at a higher delay, which requires the transmitter to detune slower at lower $f$ = 7.4 kHz, yet at a more accurate setting in terms of compatible SFR. However, for co-existing FR in parallel, as indicated for both FR at reaches of 4.3 and 7 km, a common sweep frequency has to be applied (κ). Figure 2c summarizes the SFR to realize a low spectral overlap of 1/32. It is restricted by the SFR of the lower sweep frequency $f$, which in this case is given by the second FR at its reach of 7 km. Although the tolerable SFR does not pose a technical challenge for frequency synthesizers, it shows that the overlap of SFRs may quickly restrict the values of compatible sweep frequencies (κ) in case of multiple FRs present at the ODN. A high frequency deviation ΔF for the optical emission is beneficial to find a common sweep frequency in case of multiple FR.

The presented analysis assumes a sawtooth as sweep function, for which the falling ramp is infinitely steep. It shall be stressed that this part of the sequence that resets the wavelength sweep unavoidably causes overlap with the crosstalk. By keeping the sawtooth as highly asymmetric triangular function, this overlap can be minimized. For example, a falling ramp that takes 3.13% of the period, corresponding to 1/32 ratio of the time as it also applies for a PON-oriented uplink slot in which no reception but transmission occurs, leads to an equally low overlap probability.

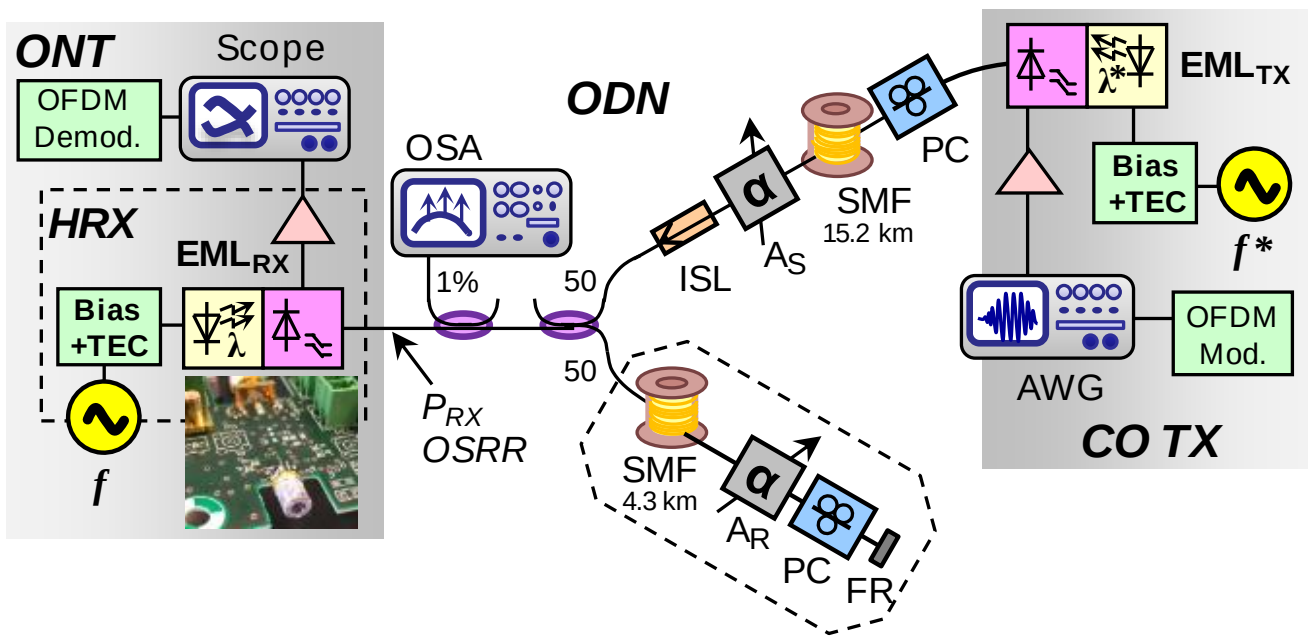

Fig. 3. Experimental setup to evaluate synchronized wavelength-swept signal transmission and its mitigation capabilities for FR.

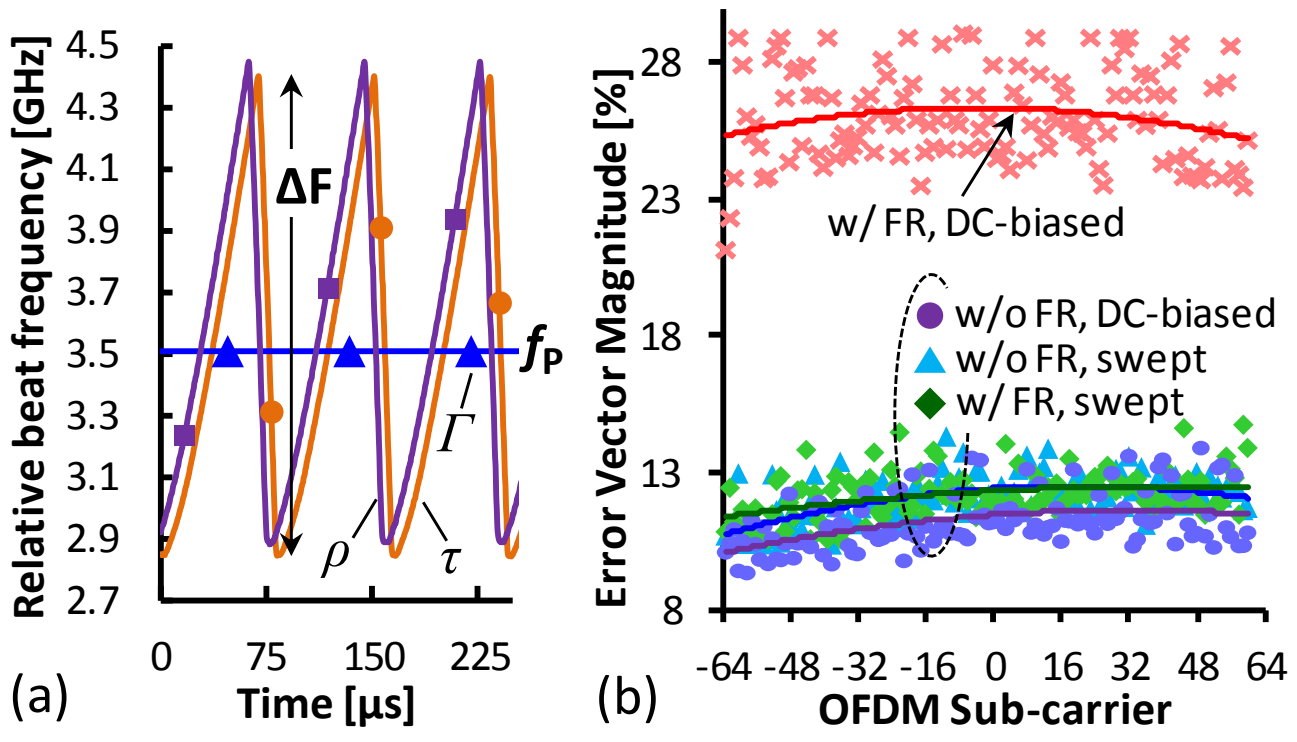

Fig. 4. (a) Beating of a pilot $f_P$ at swept (●,■) free-running $EML_{TX}$/$EML_{RX}$ emission with a stable reference and EML-to-EML transmission with synchronized sweep (▲). (b) EVM per sub-carrier at OSRR = 5 dB for reception in presence and absence of FR and mitigation technique.

The proposed concept was evaluated for coherent analogue radio-over-fiber (RoF) transmission (Fig. 3). Both, the CO transmitter and the coherent homodyne receiver (HRX) at the ONT are composed by transistor-outline EMLs operating at the channel C36 of the dense wavelength division multiplexing (DWDM) grid. Such EMLs are widely adopted and considered as cost-efficient transmitters. A 125-MHz orthogonally frequency division multiplexed (OFDM) RoF downlink was modulated on λ* and is launched with 3.5 dBm to the ODN. The fiber plant is emulated by a 15.2 km single mode fiber (SMF) and a variable optical attenuator ($A_S$). The downlink is received by the HRX through the injection-locked EML, which delivers an output power of 4.5 dBm to the ODN. An artificial FR is introduced through an unterminated FC/PC fiber patchcord at a reach of 4.3 km from the 50/50 splitter stage at the ONT, thus locating the FR inside the ODN. The optical signal-to-reflection ratio (OSRR) is set through an attenuator (AR) in the reflection branch and the polarization state of the back-reflection was set by a manual polarization controller (PC) according to a worst-case scenario. An isolator and a PC were included in the CO branch of the ODN to guarantee that the set OSRR is not influenced by further reflections and to co-polarize the LO of the $EML_{RX}$ with the downlink. Practical deployment would require a polarization-diversity HRX [13]. Full-duplex uplink transmission is omitted but has been demonstrated earlier [12].

The DFB sections of both EMLs are to be jointly modulated with a low-frequency sequence in order to synchronously sweep the emission frequencies of LO (λ) at $EML_{RX}$ and downlink seed laser (λ*) at $EML_{TX}$. Complexity-wise the same overhead as for a dithering scheme [2] applies. However, synchronization is required in terms of sweep frequency, to avoid unlocking between CO transmitter and ONT receiver, and, more importantly, for the waveform phase with a timing accuracy in the µs range. This process can build on higher-layer signaling as used for the ranging of ONTs that share a common CO receiver, but was conducted manually in this proof-of-concept work.

In a second step the same parameters for swing, waveform and frequency were chosen to ensure that the HRX remains injection-locked. Moreover, the frequency $f = f^*$ = 12.1 kHz of the asymmetric, sawtooth-like sweep was set according to the round-trip time to the FR, which is determined by the fiber length. FRs closer to the ONT requires higher sweep frequencies, which are supported by the laser's high direct modulation bandwidth in the low GHz range. Figure 4a presents the time evolution of the received pilot tone $f_P$ frequency when the tone is modulated on the free-running $EML_{TX}$ and $EML_{RX}$ output signals and these are beating with a stable reference laser (τ, ρ). There is no injection from the $EML_{TX}$ to the $EML_{RX}$. The beating signals (■,●) show a frequency modulation of the pilot tone with a deviation of ΔF = 1.55 GHz (corresponding to Λ) and therefore much larger than the 125-MHz OFDM bandwidth of the analogue RoF signal applied for later data transmission. The sweeping range thus covers spectrum out-of-band of the OFDM signal, in which crosstalk will be off-loaded. In case of wideband data signals larger ΔF values can be supported through joint temperature modulation of the EMLs. Moreover, the difference in frequency deviation among both EML outputs is 60 MHz and results from variance in the L-I characteristics and the chirp. However, it is smaller than the injection locking range. This enables proper injection locking: When beating the pilot-modulated $EML_{TX}$ output signal with the unmodulated $EML_{RX}$ signal – while exploiting the injection-locked $EML_{RX}$ as coherent receiver – the pilot frequency $f_P$ remains stable (▲) at its nominal value despite the presence of wavelength-swept operation for both EMLs (Γ). This confirms correct locking of the EML pair and thus the preservation of signal integrity.

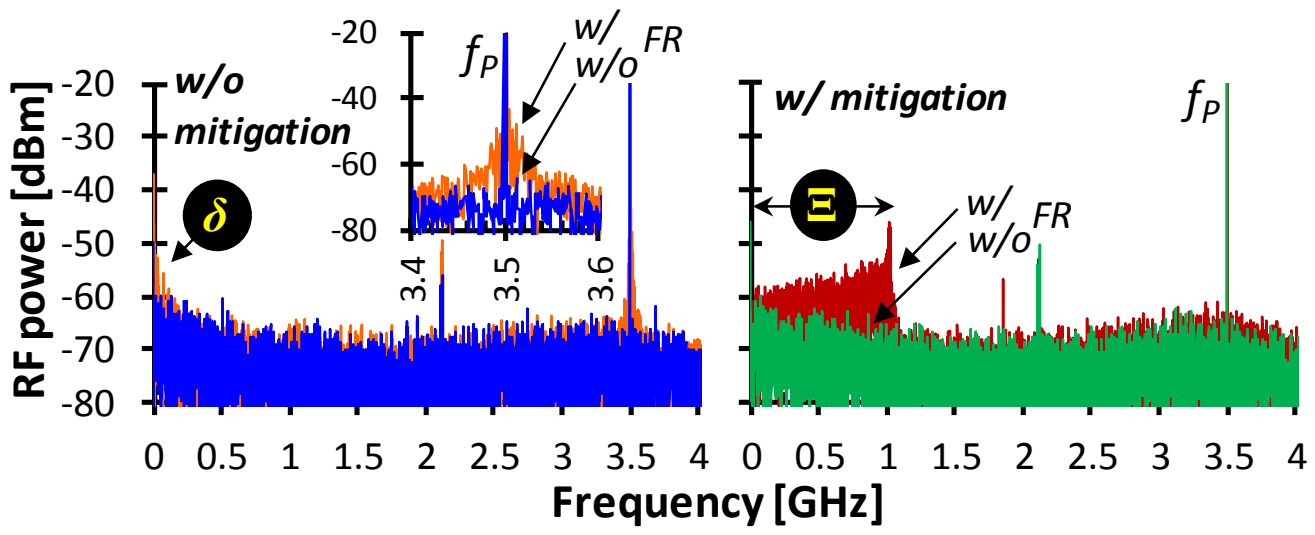

Fig. 5. Pilot tone spectrum in absence/presence of FR and under additional reflection mitigation.

The EMLs were then evaluated in the ODN with FR, thus at a reduced OSRR. Figure 5 presents the pilot tone spectra received with the $EML_{RX}$. Without FR and static DFB bias the pilot tone is clearly resembled (blue trace). This is a result of the precise photonic down-conversion to the radio frequency (RF) domain enabled by stable injection locking. In presence of the FR this locking becomes instable. A smeared beat note (δ) appears at the low frequencies as the optical carrier λ unlocks from λ*. Moreover, the pilot tone is washed out (orange) due to beat interference arising from the FR. When the DFB lasers are jointly swept, the crosstalk caused by the FR is shifted in frequency as indicated through the strong beat note deviation (Ξ, red) of the back-scattered optical LO carrier emitted by $EML_{RX}$. Crosstalk falls out-of-band to the received tone, for which the SNR is recovered. There is no degradation due to wavelength sweeping without FR (green) provided that the sweep sequences are synchronized.

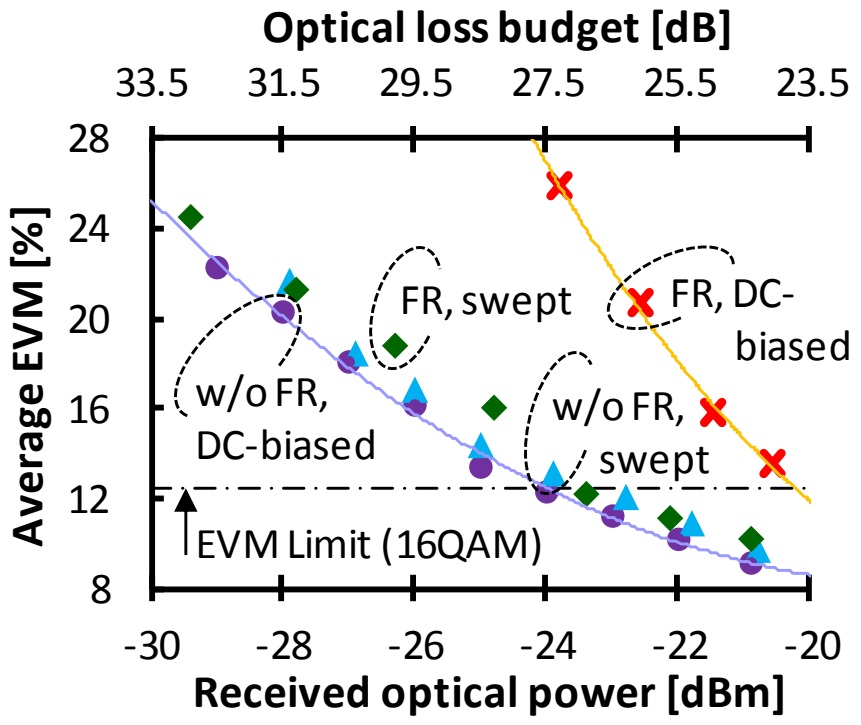

Fig. 6. EVM performance as function of the optical loss budget.

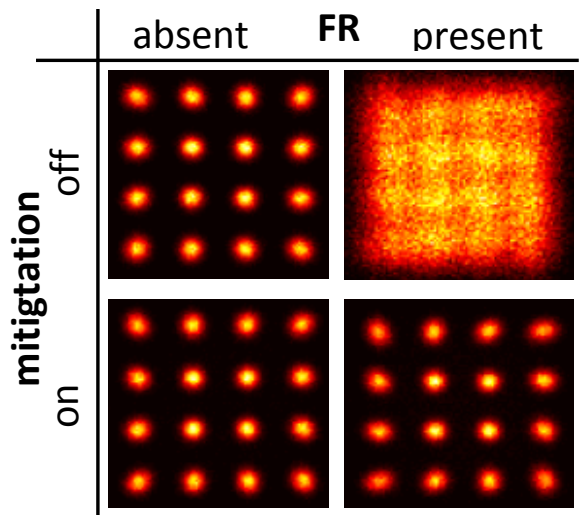

Fig. 7. Constellation diagrams for a received optical power of -22 dBm.

The reflection-tolerant operation was confirmed through measurements of the error vector magnitude (EVM) for OFDM-based RoF transmission since such analogue signals are especially vulnerable to crosstalk. Figure 4b shows the EVM for all 128 sub-carriers of the 16QAM-OFDM. In presence of the FR (×) and an OSRR of 5 dB the EVM increases by 14.8% with respect to the reflection-less reception with static laser wavelengths (●). This large penalty can be mitigated by 93% when wavelength-swept transmission is applied (◆), resulting in an average EVM of 12.2% below the 16QAM EVM limit, at an EVM penalty of 0.98%. For comparison the reflection-less yet wavelength-swept reception (▲) is shown. There is a small implementation penalty of 0.85% with respect to DC-biased light sources (●). The dependence of the EVM performance on the optical loss budget is shown in Fig. 6 for the same reflectance set earlier. For the received power range, the locking range is >200 MHz. The mitigation technique (◆) improves the loss budget by 3.6 and 4.2 dB with respect to the reflection-degraded case (×) at the EVM limits for 16QAM and QPSK, respectively. Better reception sensitivities can be obtained with an electrical EAM front-end based on a transimpedance rather than a 50Ω amplifier. Figure 7 presents the constellation diagrams for all cases for a received optical power of -22 dBm.

Figure 8 shows the EVM degradation for a fixed optical loss budget of 26.8 dB and reduced OSRR. The present FR without additional compensation (×) quickly leads to a large EVM penalty compared to the reflection-less case (●, OSRR→∞). The proposed method with swept light keeps the penalty at a value of 1.2% for the lowest OSRR of 0.8 dB (ς), where it would reach 19.4% without mitigation (ξ). This means a 94% reduction in penalty at this low OSRR value, which is characteristic for PONs: ITU-T G.989 specifies a minimum optical return loss of 32 dB, which would suggest OSRR values in the range of 0 to -10 dB for coherent PONs.

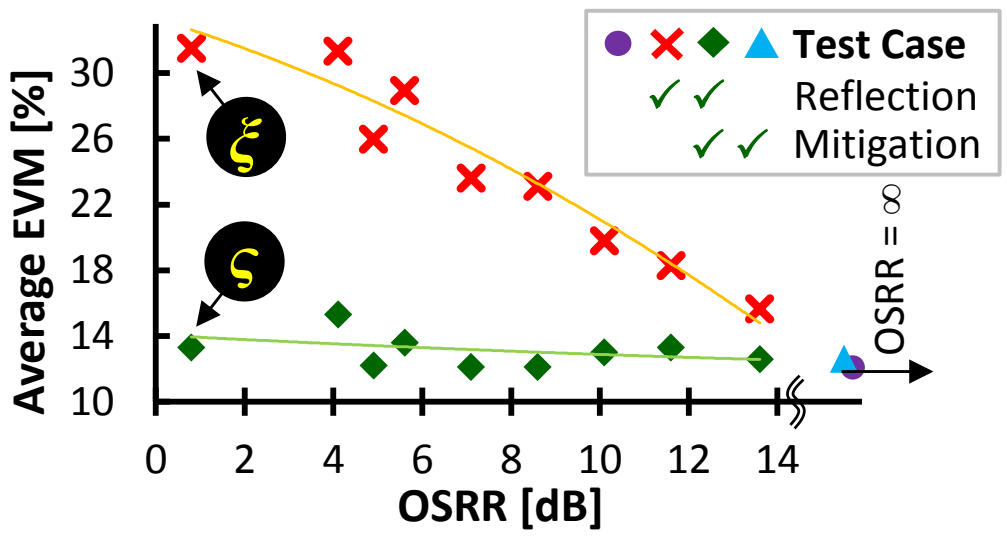

Fig. 8. EVM performance as function of the OSRR.

A mitigation scheme for reflection-induced crosstalk noise in a coherent link has been validated. By jointly sweeping the light at CO emitter and ONT receiver, crosstalk is spectrally off-loaded from the reception band. This enables operation at low OSRR of 1 dB at which the EVM penalty can be reduced by 94%. Additional temperature modulation can provide a wider sweeping range in order to accommodate higher modulation bandwidths and to potentially serve as physical-layer security feature.

## Acknowledgement

This work was supported in part by the European Research Council (ERC) under the European Union's Horizon 2020 research and innovation programme (grant agreement No 804769).